\documentstyle[aps,preprint,epsfig]{revtex}
\tightenlines
\newcommand{\bea}{\begin{eqnarray}}
\newcommand{\eea}{\end{eqnarray}}
\newcommand{\beq}{\begin{equation}}
\newcommand{\eeq}{\end{equation}}
\newcommand{\bay}{\begin{array}}
\newcommand{\eay}{\end{array}}
\begin{document}
\preprint{\parbox{6cm}{\flushright CLNS 98/1582\\TECHNION-PH-98-88\\[1cm]}}
\title{Model-independent electroweak penguins in $B$ decays to two 
pseudoscalars}
\author{Michael Gronau}
\address{Physics Department,
Technion - Israel Institute of Technology, 32000 Haifa, Israel}
\author{Dan Pirjol and Tung-Mow Yan}
\address{Floyd R. Newman Laboratory of Nuclear
Studies, Cornell University, Ithaca, New York 14853}
\date{\today}
\maketitle

\begin{abstract} 
We study the effects of electroweak penguin (EWP) amplitudes in $B$ meson decays
into two charmless pseudoscalars in the approximation of retaining only the 
dominant EWP operators $Q_9$ and $Q_{10}$. Using flavor SU(3) symmetry, we 
derive a set of model-independent relations between EWP contributions and 
tree-level decay amplitudes one of which was noted recently by Neubert and 
Rosner. Two new applications of these relations are demonstrated in which 
uncertainties due to EWP corrections are eliminated in order to determine a weak 
phase. Whereas the weak angle $\alpha$ can be obtained from $B\to\pi\pi$ free 
of hadronic 
uncertainties, a determination of $\gamma$ from $B^{0,\pm}\to K\pi^{\pm}$
requires the knowledge of a ratio of certain tree-level hadronic matrix 
elements.
The smallness of this ratio implies a useful constraint on $\gamma$ if 
rescattering can be neglected.
\end{abstract}
  
\pacs{pacs1,pacs2,pacs3}

\narrowtext

\section{Introduction}

Nonleptonic weak decays of $B$ mesons into two charmless pseudoscalars
provide an important probe of the origin of CP violation in the single complex 
phase of the CKM matrix \cite{MG}. Approximate flavor symmetries of the strong 
interactions play a useful role in such analysis \cite{Zep,GHLR,GrLe}. In 
one simplified version of such methods the weak phase $\alpha$ is extracted 
from $B \to \pi\pi$ decays using isospin symmetry \cite{GL}, and in another case
the phase $\gamma$ is obtained from combining $B\to K\pi$ and $B\to \pi\pi$ 
amplitudes using flavor SU(3) \cite{GRL}. Electroweak penguin (EWP) contributions 
\cite{Fle}, enhanced by the heavy top quark, can spoil such methods. Whereas 
these contributions are expected to have a small effect on $\alpha$, they were 
estimated in a model-dependent manner to have a large effect on the extraction 
of $\gamma$ \cite{Desh,EWP,Fleis}. Recently Neubert and Rosner have used Fierz 
transformations and SU(3) symmetry to include in the latter case the effect of 
EWP amplitudes 
in a model-independent way \cite{NR1,NR2}. Their method of constraining $\gamma$ is 
based on assuming the dominance of two EWP operators ($Q_9$ and $Q_{10}$) and 
relating their matrix elements for the $I=3/2~K\pi$ $B$ decay final state to 
corresponding tree-level amplitudes. This argument is entirely 
model-independent, in contrast to previous studies of EWP contributions 
\cite{Desh,Fleis,FL} which assume certain models for the matrix elements of 
EWP operators involving factorization and specific form factors.

The purpose of this paper is to generalize the relation proposed by Neubert and 
Rosner to all matrix elements of EWP operators for nonstrange and 
strange $B$ mesons and for any two pseudoscalar final state, and to study the 
consequences of such relations. Sec. II reviews the two alternative 
descriptions of flavor SU(3), in terms of operator matrix elements on the 
one hand, and quark diagrams on the other hand. These descriptions are used in 
Sec. III to derive a complete set of model-independent SU(3) relations between 
EWP and tree amplitudes 
for $B\to K\pi,~B\to\pi\pi,~B\to K \overline{K}$ and corresponding $B_s$ decays. 
Using an approximate numerical relation between two ratios of Wilson 
coefficients, we show in Sec. IV that {\it all} EWP 
contributions can be written in terms of tree amplitudes. In Sec. V we 
demonstrate a few applications of these relations used to eliminate 
uncertainties 
due to EWP contributions when determining the weak phases $\alpha$ and $\gamma$ 
from $B\to \pi\pi$ and $B\to K\pi$ decays, respectively. Finally, our results 
are summarized in Sec. VI. An Appendix lists the four-quark operators appearing
in the weak Hamiltonian for $b$ decays corresponding to specific SU(3) 
representations. 
 
\section{Flavor SU(3) in $B$ decays}

The weak Hamiltonian governing $B$ meson decays is given by (see, e.g.,
\cite{BBL})
\beq\label{H}
{\cal H} = \frac{G_F}{\sqrt2}
\sum_{q=d,s}\left(\sum_{q'=u,c} \lambda_{q'}^{(q)}
[c_1 (\bar bq')_{V-A}(\bar q'q)_{V-A} + c_2 (\bar bq)_{V-A}(\bar q'q')_{V-A}]
-  \lambda_t^{(q)}\sum_{i=3}^{10}c_i Q^{(q)}_i\right)~,
\eeq
where $\lambda_{q'}^{(q)}=V_{q'b}^*V_{q'q},~q=d,s,~q'=u,c,t$. Unitarity 
of the CKM matrix implies $\lambda_u^{(q)}+\lambda_c^{(q)}+
\lambda_t^{(q)}=0$. The first term, involving the coefficients $c_1$ and $c_2$, 
will be referred to as the ``tree'' part, while the second term, involving 
$c_i,~i=3-10$ is the penguin part. The corresponding $Q_i$ consist of four QCD 
penguin operators ($i=3-6$) and four electroweak penguin operators ($i=7-10$). 
Their precise form is not important for our purpose and can be found for 
example in \cite{BBL}. In the following we will be only 
interested in their SU(3) transformation properties, noting that $Q_9$ and 
$Q_{10}$ have a $(V-A)(
V-A)$ structure similar to the ``tree'' part. There are 
two 
distinct types of QCD penguin operators, with the flavor structure ($q=d,s$)
\bea\label{P}
Q^{(q)}_{3,5} &=& (\bar bq)(\bar uu+\bar dd +\bar ss)~,\nonumber\\
Q^{(q)}_{4,6} &=& (\bar bu)(\bar uq)+(\bar bd)(\bar dq)+(\bar bs)(\bar sq)~,
\eea
and two types of EWP operators
\bea\label{EWPen}
Q^{(q)}_{7,9} &=& \frac32\left[
(\bar bq)(\frac23\bar uu-\frac13\bar dd -\frac13\bar ss)\right]~,\nonumber\\
Q^{(q)}_{8,10} &=& \frac32\left[
\frac23(\bar bu)(\bar uq)-\frac13(\bar bd)(\bar dq)-\frac13(\bar bs)(\bar 
sq)\right]~.
\eea

All four quark operators appearing in (\ref{H}-\ref{EWPen}) are of the form
$(\bar bq_1)(\bar q_2 q_3)$ and can be written as a sum of ${\overline 
{\bf 15}}$, {\bf 6} and ${\bar {\bf 3}}$, into which the product ${\bar 
{\bf 3}}\otimes {\bf 3}\otimes {\bar {\bf 3}}$ can be decomposed 
\cite{GHLR,GrLe}. Note that the representation ${\bar {\bf 3}}$ appears twice 
in this decomposition, both symmetric (${\bar {\bf 3}}^{(s)}$), 
and antisymmetric (${\bar {\bf 3}}^{(a)}$) under the interchange of $q_1$ and
$q_3$.

The tree part of the Hamiltonian (\ref{H}) can be expressed in terms of 
operators with definite SU(3) transformation properties:
\bea\label{T}
{\cal H}_T &=& \frac{G_F}{\sqrt2}\left(\lambda_u^{(s)}
[\frac12(c_1-c_2)(-\bar {\bf 3}^{(a)}_{I=0} - {\bf 6}_{I=1}) +
\frac12(c_1+c_2)(-\overline {\bf 15}_{I=1} - \frac{1}{\sqrt2}\overline {\bf 15}_{I=0}
+\frac{1}{\sqrt2}\bar {\bf 3}^{(s)}_{I=0}) \right.\nonumber\\
&+& \left. \lambda_u^{(d)}
[\frac12(c_1-c_2)({\bf 6}_{I=\frac12} - \bar {\bf 3}^{(a)}_{I=\frac12}) +
\frac12(c_1+c_2)(-\frac{2}{\sqrt3}\overline {\bf 15}_{I=\frac32} - 
\frac{1}{\sqrt6}\overline {\bf 15}_{I=\frac12}
+\frac{1}{\sqrt2}\bar {\bf 3}^{(s)}_{I=\frac12}) \right)~.
\eea
The operators ${\bar {\bf 3}}^{(s)}$ and 
${\bar {\bf 3}}^{(a)}$ appear in the two lines in the same combination.
This fact is essential for relating $|\Delta S|=1$ to $\Delta S=0$
amplitudes with the help of SU(3) symmetry. The operators with well-defined 
SU(3) transformation properties appearing in (\ref{T}) are given in 
the Appendix in terms of four-quark operators.

The contribution of the EWP operators (\ref{EWPen}) is given by
\bea\label{EWP}
{\cal H}_{EWP} &\simeq & -\lambda_t^{(s)}\left(c_9 Q^{(s)}_9 + c_{10} 
Q^{(s)}_{10}\right) -
\lambda_t^{(d)}\left(c_9 Q^{(d)}_9 + c_{10} Q^{(d)}_{10}\right) = \\
&-& \frac{\lambda_t^{(s)}}{2}\left(
\frac{c_9-c_{10}}{2}(3\cdot {\bf 6}_{I=1} + \bar {\bf 3}^{(a)}_{I=0} ) +
\frac{c_9+c_{10}}{2}( -3\cdot\overline {\bf 15}_{I=1} 
-\frac{3}{\sqrt2}\overline {\bf 15}_{I=0}
-\frac{1}{\sqrt2}\bar {\bf 3}^{(s)}_{I=0} )\right)\nonumber\\
& &\hspace{-2cm} -\frac{\lambda_t^{(d)}}{2}\left(
\frac{c_9-c_{10}}{2}(-3\cdot {\bf 6}_{I=\frac12} + \bar {\bf 3}^{(a)}_{I=\frac12} ) +
\frac{c_9+c_{10}}{2}( -\sqrt{\frac32}\cdot\overline {\bf 15}_{I=\frac12} 
-2\sqrt3\cdot \overline {\bf 15}_{I=\frac32}
-\frac{1}{\sqrt2}\bar {\bf 3}^{(s)}_{I=\frac12} )\right)~,\nonumber
\eea
where we made the approximation of keeping only contributions from
$Q_9$ and $Q_{10}$ \cite{Fle,NR1}. This is justified by the tiny Wilson 
coefficients of the remaining two operators $Q_7$ and $Q_8$ \cite{BBL}.
In this approximation the operators appearing in (\ref{EWP}) are of the 
$(V-A)(V-A)$ type and can be related to those 
appearing in the tree Hamiltonian (\ref{T}). It is this fact which will allow 
us to express EWP contributions in terms of tree-level decay amplitudes.

Before proceeding to obtain these relations, let us recall the equivalent 
description of SU(3) amplitudes in terms of quark diagrams \cite{GHLR}. There 
are six topologies, representing tree ($T$), color-suppressed ($C$),
annihilation ($A$), $W$-exchange ($E$), penguin ($P$) and penguin-annihilation 
($PA$) amplitudes. The six amplitudes appear in five distinct combinations,
separately for $\Delta S =0$ and $\Delta S = 1$ transitions. 
For convenience, we define these amplitudes such that they don't include the 
CKM factors. For example, a typical $|\Delta S|=1$ transition amplitude is 
\bea\label{S=1}
A(B^+\to  K^0\pi^+) = \lambda_u^{(s)}(P_u+A) + \lambda_c^{(s)}P_c +
\lambda_t^{(s)}(P_t+P^{EW}_t(B^+\to  K^0\pi^+))~,
\eea
where $P_u, A$ and $P_c$ are contributions from the four-quark operators in 
the first term of (\ref{H}), while $P_t$ and $P^{EW}_t$ originate from the 
second term. In a similar way, a typical $\Delta S=0$ transition amplitude
has the form
\bea\label{S=0}
A(B^0 \to \pi^+\pi^-) &=& \lambda_u^{(d)}(-P_u-T-E-PA_u) + \lambda_c^{(d)}(-P_c)\\
&+& \lambda_t^{(d)}(-P_t-PA_t+P^{EW}_t(B^0 \to \pi^+\pi^-))~.\nonumber
\eea
Despite their name, $P_u$ and $P_c$ originate purely from ``tree-level''
four-quark operators, . Note that in the SU(3) symmetric limit, 
the same hadronic parameters $P_u, T, C, A, PA_u, P_c, P_t$ appear in
$|\Delta S| = 1$ and  $\Delta S = 0$ transitions.

It is straightforward to relate the ``graphical'' hadronic parameters $P_u, PA_u, 
T , C, A, 
E$ to SU(3) reduced matrix elements of the operators appearing in (\ref{T}). 
This was done in the appendix of \cite{GHLR}, and can also be done by computing
representative decay amplitudes and expressing them with the help of the 
relations in the Appendix of \cite{GrLe}. We find the following set of linearly 
independent relations
\bea\label{graphic}
P_u+T &=& \frac{3}{2\sqrt{10}}a_2 + \frac12\sqrt{\frac35}a_3 + 
\frac14\sqrt{\frac35} a_4 - \frac23\sqrt{\frac25}a_5~,\nonumber\\
P_u+A &=& \frac{3}{2\sqrt{10}}a_2 - \frac12\sqrt{\frac35}a_3 - 
\frac34\sqrt{\frac35} a_4 + \frac{2}{3\sqrt{10}}a_5~,\nonumber\\
-P_u+C &=& -\frac34\sqrt{\frac25}a_2 - \frac12\sqrt{\frac35}a_3 - 
\frac14\sqrt{\frac35} a_4 - \sqrt{\frac{2}{5}}a_5~,\nonumber\\
P_u+PA_u &=& -\frac12 a_1 + \frac{1}{2\sqrt{10}}a_2 - \frac12\sqrt{\frac35}a_3 
+ \frac34\sqrt{\frac35}a_4 + \frac{1}{6\sqrt{10}}a_5~,\nonumber\\
C-E &=& -\sqrt{\frac35}a_3 + \sqrt{\frac35}a_4 - \sqrt{\frac25}a_5~.
\eea
$a_i$ denote the following combinations of reduced matrix elements 
(a factor $G_F/\sqrt2$ is omitted for simplicity)
\bea\label{ai}
a_1 &=& \frac12(c_1+c_2) \frac{1}{\sqrt2}\langle {\bf 1}|\!| {\bar 
{\bf 3}}^{(s)} |\!| {\bf 3}\rangle -
\frac12(c_1-c_2) \langle {\bf 1}|\!| {\bar {\bf 3}}^{(a)} |\!|{\bf 3}\rangle~,
\nonumber\\
a_2 &=& \frac12(c_1+c_2) \frac{1}{\sqrt2}\langle {\bf 8}|\!| {\bar 
{\bf 3}}^{(s)} |\!| {\bf 3}\rangle -
\frac12(c_1-c_2) \langle {\bf 8}|\!| {\bar {\bf 3}}^{(a)} |\!|{\bf 3}\rangle~,
\nonumber\\
a_3 &=& -\frac12(c_1-c_2) \langle {\bf 8}|\!| {\bf 6} |\!|{\bf 3}\rangle~,
\nonumber\\
a_4 &=& \frac12(c_1+c_2) \langle {\bf 8}|\!| \overline{{\bf 15}} 
|\!|{\bf 3}\rangle~,
\nonumber\\
a_5 &=& \frac12(c_1+c_2) \langle {\bf 27}|\!| \overline{{\bf 15}} 
|\!|{\bf 3}\rangle~.
\eea
The normalization of the reduced matrix elements is chosen as in \cite{GrLe}.
Relative normalization with respect to the one used in \cite{GHLR} is given in 
the Appendix.

One can find three combinations of graphical amplitudes which are independent
of the reduced matrix elements $a_1, a_2$. As explained in the next section,
they will be useful in relating EWP contributions to tree amplitudes.
\bea\label{3graph}
T-A &=& \sqrt{\frac35}a_3 + \sqrt{\frac35}a_4 - \sqrt{\frac25}a_5~,
\nonumber\\
T+C &=& - \frac{\sqrt{10}}{3} a_5~,
\nonumber\\
C-E &=& -\sqrt{\frac35}a_3 + \sqrt{\frac35}a_4 - \sqrt{\frac25}a_5~.
\eea
These relations can be solved for $a_3\,, a_4$ and $a_5$
\bea\label{3ai}
a_3 &=& -\frac12 \sqrt{\frac53}(A+C-T-E)~,
\nonumber\\
a_4 &=& \frac12 \sqrt{\frac53}(-A-\frac15 C-\frac15 T-E)~,
\nonumber\\
a_5 &=& -\frac{3}{\sqrt{10}}(T+C)~.
\eea
In Sec.~IV we will need also the results for the reduced matrix elements
$a_1$ and $a_2$ expressed in terms of graphical contributions
\bea
a_1 &=& -\frac12 T + \frac16 C - \frac43 E - \frac43 P_u - 2PA_u\nonumber\\
a_2 &=& \frac12\sqrt{\frac52}\left(T - \frac13 C + A - \frac13 E + \frac83 P_u
\right)\,.
\eea

\section{Relations between EWP and tree amplitudes}

Our purpose is to relate in the SU(3) limit EWP contributions to tree 
amplitudes. We note that the operators ${\bar {\bf 3}}^{(s)}$ and 
${\bar {\bf 3}}^{(a)}$ occur in (\ref{EWP}) in different combinations than in 
(\ref{T}).
Therefore, for arbitrary values of $c_1,~c_2,~c_9$ and $c_{10}$, symmetry 
relations
for EWP contributions can only be obtained which are independent of the matrix 
elements of ${\bar {\bf 3}}^{(s)}$ and ${\bar {\bf 3}}^{(a)}$. The respective
EWP contributions can then be expressed only in terms of tree-level amplitudes $T,C,A,E$ 
with the help of the relations (\ref{3ai}).

\subsection{$|\Delta S|=1$ amplitudes}

EWP contributions to $B\to K\pi$ decays can be easily computed using the 
Hamiltonian (\ref{EWP}). One obtains
\bea\label{BEWP}
P^{EW}(B^0\to K^+\pi^-) &=& \frac{3}{4\sqrt{10}}b_2 + \frac14\sqrt{\frac35}b_3
+ \frac38\sqrt{\frac35} b_4 - \sqrt{\frac25} b_5~,
\nonumber\\
P^{EW}(B^+\to K^0\pi^+) &=& -\frac{3}{4\sqrt{10}}b_2 + \frac14\sqrt{\frac35}b_3
+ \frac98\sqrt{\frac35} b_4 - \frac{1}{\sqrt{10}} b_5~,
\nonumber\\
P^{EW}(B^0\to K^0\pi^0) &=& -\frac{3}{8\sqrt{5}}b_2 - \frac14\sqrt{\frac{3}{10}}b_3
- \frac38\sqrt{\frac{3}{10}} b_4 - \frac{3}{2\sqrt5} b_5~,
\nonumber\\
P^{EW}(B^+\to K^+\pi^0) &=& \frac{3}{8\sqrt{5}}b_2 - 
\frac14\sqrt{\frac{3}{10}}b_3 - \frac98\sqrt{\frac{3}{10}} b_4 - 
\frac{2}{\sqrt5} b_5~.
\eea
The parameters $b_i$, analogous to $a_i$, are defined as
\bea\label{bi}
b_1 &=& 
-\frac12(c_9+c_{10}) \frac{1}{\sqrt2}\langle {\bf 1}|\!| {\bar {\bf 3}}^{(s)} 
|\!| {\bf 3}\rangle + \frac12(c_9-c_{10}) \langle {\bf 1}|\!| {\bar 
{\bf 3}}^{(a)} |\!|{\bf 3}\rangle~,
\nonumber\\
b_2 &=& -\frac12(c_9+c_{10}) \frac{1}{\sqrt2}\langle {\bf 8}|\!| {\bar 
{\bf 3}}^{(s)} |\!|{\bf 3}\rangle + \frac12(c_9-c_{10}) \langle {\bf 8}|\!| 
{\bar {\bf 3}}^{(a)} |\!|{\bf 3}\rangle~,
\nonumber\\
b_3 &=& \frac32(c_9-c_{10})\langle {\bf 8}|\!| {\bf 6} |\!|{\bf 3}\rangle~,
\nonumber\\
b_4 &=& \frac12(c_9+c_{10}) \langle {\bf 8}|\!| \overline{{\bf 15}} 
|\!|{\bf 3}\rangle~,
\nonumber\\
b_5 &=& \frac12(c_9+c_{10}) \langle {\bf 27}|\!| \overline{{\bf 15}} 
|\!|{\bf 3}\rangle~.
\eea
The EWP contributions satisfy the isospin relation (as do the full amplitudes
\cite{NQ})
\bea\label{isospin}
& &P^{EW}(B^+\to K^0\pi^+) + \sqrt2 P^{EW}(B^+\to K^+\pi^0) =
\nonumber\\
& &\qquad\qquad \sqrt2 P^{EW}(B^0\to K^0\pi^0) + P^{EW}(B^0\to K^+\pi^-)~.
\eea

It is clear now that any combination of $P^{EW}$ amplitudes which is independent
of $b_1, b_2$ can be expressed directly in terms of the tree-level amplitudes 
$T, C, A, E$ using the relations (\ref{3ai})
\bea
b_3 &=& -3\frac{c_9-c_{10}}{c_1-c_2} a_3 = \frac{c_9-c_{10}}{c_1-c_2}\,
\frac{\sqrt{15}}{2}(A+C-T-E)~,
\nonumber\\
b_4 &=& \frac{c_9+c_{10}}{c_1+c_2} a_4 = \frac12\sqrt{\frac53}
\frac{c_9+c_{10}}{c_1+c_2}(-A-\frac15 C-\frac15 T-E)~,
\nonumber\\
b_5 &=& \frac{c_9+c_{10}}{c_1+c_2} a_5 = 
-\frac{3}{\sqrt{10}}\frac{c_9+c_{10}}{c_1+c_2}(T+C)~.
\eea

One can form two combinations of electroweak penguin contributions in 
$B\to K\pi$ decays which do not depend on $b_1, b_2$:
\bea\label{EW1}
& &P^{EW}(B^+\to K^0\pi^+) + \sqrt2 P^{EW}(B^+\to K^+\pi^0) = -\sqrt{\frac52} 
b_5 = \frac32\frac{c_9+c_{10}}{c_1+c_2} (T+C)~,\\\label{EW2}
& &P^{EW}(B^0\to K^+\pi^-) + P^{EW}(B^+\to K^0\pi^+) =
\frac12\sqrt{\frac35}b_3 +\frac32 \sqrt{\frac35}b_4 - 
\frac32\sqrt{\frac25}b_5
\nonumber\\
& &\qquad\qquad =\frac34\frac{c_9-c_{10}}{c_1-c_2} (A+C-T-E) -
\frac34\frac{c_9+c_{10}}{c_1+c_2} (A-C-T+E)~.
\eea
A third combination $P^{EW}(B^0\to K^0\pi^0) + P^{EW}(B^+\to K^+\pi^0)$ is not
independent of these two in view of the isospin identity (\ref{isospin}).
The first relation (\ref{EW1}) was obtained in \cite{NR1}. The second one 
(\ref{EW2}) is new.

In a similar way one can compute EWP contributions to $B_s$ decay
amplitudes. We find
\bea\label{BsEWP}
P^{EW}(B_s\to\pi^+ \pi^-) &=& -\frac14 b_1 - \frac{1}{2\sqrt{10}} b_2 - 
\frac34\sqrt{\frac35} b_4 + \frac{1}{4\sqrt{10}} b_5~,
\nonumber\\
P^{EW}(B_s\to\pi^0 \pi^0) &=& \frac{1}{4\sqrt2} b_1 + \frac{1}{4\sqrt5} b_2 + 
\frac34\sqrt{\frac{3}{10}} b_4 - \frac{1}{8\sqrt5} b_5~,
\nonumber\\
P^{EW}(B_s\to K^+ K^-) &=& -\frac14 b_1 + \frac{1}{4\sqrt{10}} b_2 + 
\frac14\sqrt{\frac35} b_3 - \frac38\sqrt{\frac35} b_4 - \frac{7}{4\sqrt{10}} 
b_5~,
\nonumber\\
P^{EW}(B_s\to K^0 \bar K^0) &=& \frac14 b_1 - \frac{1}{4\sqrt{10}} b_2 +
\frac14\sqrt{\frac35} b_3 - \frac98\sqrt{\frac35} b_4 - \frac{1}{4\sqrt{10}} 
b_5~.
\eea

Eliminating $b_1, b_2$ gives two relations
\bea\label{BsPP}
& &P^{EW}(B_s\to\pi^+ \pi^-) + \sqrt2 P^{EW}(B_s\to\pi^0 \pi^0) = 0~,\\
\nonumber
& &P^{EW}(B_s\to K^+ K^-) + P^{EW}(B_s\to K^0 \bar K^0)\\
& &\qquad\qquad  = \frac34
\frac{c_9-c_{10}}{c_1-c_2}(A+C-T-E) + 
\frac34\frac{c_9+c_{10}}{c_1+c_2}(A+C+T+E)~.
\eea
The first relation is simply a consequence of the absence of $\Delta I=2$
terms in the EWP Hamiltonian (\ref{EWP}).

\subsection{$\Delta S=0$ amplitudes}

For this case the Hamiltonian (\ref{EWP}) gives the following results for $B$
and $B_s$ decays
\bea\label{S0EWP}
P^{EW}(B^+\to \pi^+ \pi^0) &=& -\frac{\sqrt5}{2}b_5~,
\nonumber\\
P^{EW}(B^0\to \pi^+ \pi^-) &=& -\frac14 b_1 + \frac{1}{4\sqrt{10}} b_2 + 
\frac14\sqrt{\frac35} b_3 - \frac38\sqrt{\frac35} b_4 - \frac{7}{4\sqrt{10}} 
b_5~,
\nonumber\\
P^{EW}(B^0\to \pi^0 \pi^0) &=& \frac{1}{4\sqrt2} b_1 - \frac{1}{8\sqrt{5}} b_2 - 
\frac14\sqrt{\frac{3}{10}} b_3 + \frac38\sqrt{\frac{3}{10}} b_4 - 
\frac{13}{8\sqrt{5}} b_5~,
\nonumber\\
P^{EW}(B^+\to K^+ \bar K^0) &=&  - \frac{3}{4\sqrt{10}} b_2 +
\frac14\sqrt{\frac35} b_3 + \frac98\sqrt{\frac35} b_4 - \frac{1}{\sqrt{10}} 
b_5~,
\nonumber\\
P^{EW}(B^0\to K^+  K^-) &=&  -\frac14 b_1 - \frac{1}{2\sqrt{10}} b_2 -
\frac34\sqrt{\frac35} b_4 + \frac{1}{4\sqrt{10}} b_5~,
\nonumber\\
P^{EW}(B^0\to K^0 \bar K^0) &=&  \frac14 b_1 - \frac{1}{4\sqrt{10}} b_2 +
\frac14\sqrt{\frac35} b_3 - \frac98\sqrt{\frac35} b_4 - \frac{1}{4\sqrt{10}} 
b_5~,
\nonumber\\
P^{EW}(B_s\to K^-\pi^+) &=& \frac{3}{4\sqrt{10}} b_2 + 
\frac14\sqrt{\frac35} b_3 + \frac38\sqrt{\frac35} b_4 - \sqrt{\frac25} b_5~,
\nonumber\\
P^{EW}(B_s\to \bar K^0\pi^0) &=& - \frac{3}{8\sqrt{5}} b_2 - 
\frac14\sqrt{\frac{3}{10}} b_3 - \frac38\sqrt{\frac{3}{10}} b_4 - 
\frac{3}{2\sqrt{5}} b_5~.
\eea

Eliminating $b_{1-4}$ gives the following relations for EWP contributions to 
$B\to\pi\pi$ decays
\bea
\sqrt2 P^{EW}(B^+\to \pi^+ \pi^0) &=& 
P^{EW}(B^0\to \pi^+ \pi^-) + \sqrt2 P^{EW}(B^0\to \pi^0 \pi^0)\nonumber\\
 &=& \frac32\frac{c_9+c_{10}}{c_1+c_2}(T+C)~.\label{Bpipi}
\eea
This relation, describing decay amplitudes into two pions in 
a $I=2$ state, follows from isospin alone. Only the $\Delta I=3/2$ part of the 
Hamiltonian contributes to these amplitudes. Comparing the tree-level 
(\ref{T}) and EWP (\ref{EWP}) Hamiltonians, one observes that their 
$\Delta I=3/2$ parts are simply related by
\bea
{\cal H}^{EW}_{\Delta I=3/2} = -\frac32 \frac{\lambda_t^{(d)}}{\lambda_u^{(d)}}
\frac{c_9+c_{10}}{c_1+c_2} {\cal H}^{tree}_{\Delta I=3/2}~.
\eea
Therefore isospin symmetry alone suffices to relate their matrix elements. 
A similar relation holds for EWP contribution in $B_s\to 
(\overline{K}\pi)_{I=3/2}$
\bea
P^{EW}(B_s\to K^-\pi^+) + \sqrt2 P^{EW}(B_s\to \bar K^0\pi^0) =
\frac32\frac{c_9+c_{10}}{c_1+c_2}(T+C)~.
\eea

\section{Graphical representation for EWP}

The numerical values of the two ratios of Wilson coefficients appearing in
the previous section are very close to each other
\bea\label{ratios}
\frac{c_9+c_{10}}{c_1+c_2} = -1.139\alpha~,\qquad
\frac{c_9-c_{10}}{c_1-c_2} = -1.107\alpha~.
\eea
We used here the leading log values of the Wilson coefficients at $m_b$ 
\cite{BBL}
\bea
c_1=1.144~,\quad c_2=-0.308~,\quad
c_9=-1.280\alpha~,\quad c_{10}=0.328\alpha~,
\eea
with $\alpha=1/129$. The two values in (\ref{ratios})
differ by less that $3\%$. Therefore, they can be taken as having a common 
value to a very good approximation
\bea\label{approx}
\frac{c_9+c_{10}}{c_1+c_2} = \frac{c_9-c_{10}}{c_1-c_2} = \kappa~,
\eea
where $\kappa \simeq -1.123\alpha$. As a consequence of this approximate 
equality, all EWP reduced matrix elements (\ref{bi}) are proportional to 
the corresponding tree amplitudes (\ref{ai}) with a common proportionality 
constant
\bea\label{ab}
b_1 = -\kappa a_1~,\quad
b_2 = -\kappa a_2~,\quad
b_3 = -3\kappa a_3~,\quad
b_4 = \kappa a_4~,\quad
b_5 = \kappa a_5~.
\eea

These equalities suggest introducing the following six EWP amplitudes, 
analogous to the ones used to parametrize tree-level decay amplitudes
\bea\label{proport}
P^{EW}_i = \kappa i~,\qquad i=T,C,A,E,P_u,PA_u~.
\eea
These amplitudes have a direct graphic interpretation in terms of quark diagrams
with one insertion of an electroweak penguin operator.
Furthermore, the simple proportionality relation (\ref{proport}) guarantees that the 
$P^{EW}_i$ amplitudes will satisfy the same hierarchy of sizes as the 
tree-level amplitudes \cite{GHLR,EWP}.

\begin{center}
\begin{tabular}{r|cccccc}
\hline
\hline
Decay mode   & $P^{EW}_T$  & $P^{EW}_C$ & $P^{EW}_A$ & $P^{EW}_E$ & 
   $P^{EW}_{P_u}$ & $P^{EW}_{PA_u}$ \\
\hline
$B^+\to \pi^+\pi^0$ & $3/2\sqrt2$ & $3/2\sqrt2$ & 0 & 0 & 0 & 0 \\
$K^+\bar K^0$ & 0 & $1/2$ & 0 & $-1$ & $1/2$ & 0 \\
 &  & & & & &  \\
$B^0\to \pi^+\pi^-$ & 0 & 1 & 1/2 & $-1/2$ & $-1/2$ & $-1/2$ \\
$\pi^0 \pi^0$ & $3/2\sqrt2$ & $1/2\sqrt2$ & $-1/2\sqrt2$ & $1/2\sqrt2$ & 
   $1/2\sqrt2$ & $1/2\sqrt2$ \\
$K^+ K^-$ & 0 & 0 & 1/2 & 0 & 0 & $-1/2$ \\
$K^0 \bar K^0$ & 0 & $1/2$ & 1 & $1/2$ & $1/2$ & $1/2$ \\
 &  & & & & &  \\
$B_s\to K^-\pi^+$ & 0 & 1 & 0 & $-1/2$ & $-1/2$ & 0 \\
$\bar K^0\pi^0$ & $3/2\sqrt2$ & $1/2\sqrt2$ & 0 & $1/2\sqrt2$ & 
   $1/2\sqrt2$ & 0 \\
\hline
\hline
\end{tabular}
\end{center}
\begin{quote} {\bf Table 1.} EW penguin contributions to $\Delta S=0$ 
transitions in terms of the graphical amplitudes $P^{EW}_i$.
\end{quote}

Inserting the relations (\ref{ab}) into (\ref{graphic}) one may express the
parameters $b_i$ in terms of $P^{EW}_i$. Using (\ref{BEWP}), (\ref{BsEWP})
and (\ref{S0EWP}), EWP contributions to any given decay can be written as a 
linear combination of the $P^{EW}_i$ amplitudes. The results are given 
in Table 1 for $\Delta S=0$ transitions and in Table 2 for $|\Delta S|=1$ 
decays.

\begin{center}
\begin{tabular}{r|cccccc}
\hline
\hline
Decay mode   & $P^{EW}_T$  & $P^{EW}_C$ & $P^{EW}_A$ & $P^{EW}_E$ & 
   $P^{EW}_{P_u}$ & $P^{EW}_{PA_u}$ \\
\hline
$B^+\to K^0\pi^+$ & 0 &  $1/2$ & 0 & $-1$ & $1/2$ & 0 \\
$K^+\pi^0$ & $3/2\sqrt2$ & $1/\sqrt2$ & 0 & $1/\sqrt2$ & 
   $-1/2\sqrt2$ & 0 \\
 &  & & & & &  \\
$B^0\to K^+\pi^-$ & 0 & 1 & 0 & $-1/2$ & $-1/2$ & 0 \\
$K^0\pi^0$ & $3/2\sqrt2$ & $1/2\sqrt2$ & 0 & $1/2\sqrt2$ & 
   $1/2\sqrt2$ & 0 \\
 &  & & & & &  \\
$B_s\to \pi^+\pi^-$ & 0 & 0 & 1/2 & 0 & 0 & $-1/2$ \\
$\pi^0\pi^0$ & 0 & 0 & $-1/2\sqrt2$ & 0 & 
   0 & $1/2\sqrt2$ \\
$K^+ K^-$ & 0 & 1 & 1/2 & $-1/2$ & 
   $-1/2$ & $-1/2$ \\
$K^0 \bar K^0$ & 0 & $1/2$ & 1 & $1/2$ & 
   $1/2$ & $1/2$ \\
\hline
\hline
\end{tabular}
\end{center}
\begin{quote} {\bf Table 2.} EW penguin contributions to $|\Delta S|=1$ 
transitions in terms of the graphical amplitudes $P^{EW}_i$.
\end{quote}

The results in Tables 1 and 2 agree with a previous analysis of the EWP
contributions in quark diagram language \cite{EWP}. The relation between the 
EWP amplitudes of \cite{EWP} and our parameters $P^{EW}_i$ is given by
\bea
P_{EW} &=& -\frac32 \lambda_t^{(d)} P^{EW}_T~,\qquad 
P_{EW}^C = -\frac32 \lambda_t^{(d)} P^{EW}_C~,
\nonumber\\
P'_{EW} &=& -\frac32 \lambda_t^{(s)} P^{EW}_T~,\qquad 
P_{EW}^{'C} = -\frac32 \lambda_t^{(s)} P^{EW}_C~.
\eea
The improvement over \cite{EWP} is that these parameters can be simply expressed
through (\ref{proport}) in terms of tree-level graphical amplitudes. Thus, the 
effects of EWP contributions can be included to a good approximation in a 
model-independent way without encountering any new hadronic amplitudes. 
One of the consequences of this simplification is that color-suppression of
certain EWP amplitudes is identical to the corresponding suppression of tree 
amplitudes, and does not require further assumptions about hadronic matrix 
elements of EWP operators.

\section{Applications}

\subsection{Determination of $\alpha$ from $B\to\pi\pi$ decays }

It has been proposed in \cite{GL} to determine the weak angle $\alpha$
from a combined measurement of the time-dependent decay rate
$B^0(t)\to\pi^+\pi^-$ and time-integrated branching ratios for $B^+\to\pi^+\pi^0$,
$B^0\to\pi^+\pi^-$, $B^0\to\pi^0\pi^0$ and their CP-conjugated modes.
As noted in \cite{Desh,EWP,Fleis}, this method is affected by uncertainties 
arising from the presence of EWP contributions. We will show in the following 
how their effect can be taken into account in a model-independent way 
\cite{BuFl}.

The angle $\alpha$ is measured through the time-dependent decay rate $B^0(t)\to
\pi^+\pi^-$ which contains a term of the form
\bea
|\langle\pi^+\pi^-|B^0(t)\rangle |^2 = \cdots + 
|A(B^0\to \pi^+\pi^-)|\, |A(\bar B^0\to \pi^+\pi^-)|e^{-\Gamma t}
\sin(2\alpha + \theta)\sin (\Delta m t)~,
\eea
$\Delta m$ being the mass difference between the two neutral $B$ mass 
eigenstates. The angle $\theta$ is due to the presence of QCD penguins in the
$B^0\to \pi^+\pi^-$ amplitude and is defined as
$\theta = \mbox{Arg}(\tilde A(\bar B^0\to \pi^+\pi^-)/A(B^0\to \pi^+\pi^-))$
(with $\tilde A(\bar B\to\bar f) \equiv e^{2i\gamma}A(\bar B\to \bar f)$).

The idea of \cite{GL} is to measure $\theta$ through a geometrical construction.
An essential ingredient of the method is the equality of the following
two decay amplitudes
\beq\label{equality}
A(B^+\to\pi^0\pi^+) = \tilde A(B^-\to\pi^0\pi^-)~,
\eeq
which can be therefore taken as the common base of two isospin triangles
for the decays $B^+\to\pi^+\pi^0$, $B^0\to\pi^+\pi^-$, $B^0\to\pi^0\pi^0$
and their CP-conjugate modes. The angle $\theta$ is obtained from this 
construction as $\theta=$Arg$(\tilde A(\bar B^0\to \pi^+\pi^-)/A(B^0\to \pi^+\pi^-))$.

The equality (\ref{equality}) is spoiled in the presence
of the EWP terms, in which case one has
\bea\label{B+pi0pi+}
\sqrt2 A(B^+\to\pi^0\pi^+) = -\lambda_u^{(d)}(T+C) + \lambda_t^{(d)}\frac32
\frac{c_9+c_{10}}{c_1+c_2}(T+C)~.
\eea
We made use of the isospin relation (\ref{Bpipi}) for the EWP contribution 
to this decay.

The amplitude (\ref{B+pi0pi+}) and its CP-conjugate are shown in Figure 1, from 
which
two conclusions are immediately apparent: a) the equality between the decay rates
for $B^+\to\pi^0\pi^+$ and its CP-conjugate holds also in the presence of the
EWP amplitudes; b) the value of the angle $2\xi$ between $A(B^+\to\pi^0\pi^+)$ 
and $\tilde A(B^-\to\pi^0\pi^-)$ is a calculable function of $\alpha$ alone.
A simple calculation gives
\bea
\tan\xi = \frac{x\sin\alpha}{1+x\cos\alpha}~,\qquad
x\equiv \frac32\frac{c_9+c_{10}}{c_1+c_2}
\left|\frac{\lambda_t^{(d)}}{\lambda_u^{(d)}}\right| = -0.013
\left|\frac{\lambda_t^{(d)}}{\lambda_u^{(d)}}\right|~,
\eea
where $|\lambda_t^{(d)}/\lambda_u^{(d)}|=|V_{tb}V_{td}/V_{ub}V_{ud}|\approx
|V_{td}/V_{ub}|$. Note that the angle $\xi$ depends only on $\alpha$ 
and on the parameter $x$ which involves some uncertainty in its CKM factor, 
but is free of any hadronic uncertainty.

Therefore, the method proposed in \cite{GL} can be adapted to include the 
effects of the EWP by defining
the modified amplitudes $\tilde A'(\bar B\to \bar f)=e^{2i\xi}\tilde A(\bar B\to \bar f)$
in terms of which the equality (\ref{equality}) is restored.
The geometrical construction of \cite{GL} can be carried through as before and 
$\theta$ is extracted as
\bea
\theta = \mbox{Arg}\frac{\tilde A'(\bar B^0\to \pi^+\pi^-)}{A(B^0\to \pi^+\pi^-)} \mp
2\xi(\alpha)~.
\eea
The upper (lower) sign in this formula corresponds to the case when the two
triangles are drawn on the same (on opposite) side of the common amplitude 
(\ref{equality}). As in the original version of this method, there is a 
four-fold ambiguity in the value of $\alpha$, arising from the above mentioned 
freedom in the geometric construction and from having to extract $\alpha$ from 
$\sin(2\alpha+\theta)$.

Numerically the shift in the angle $\Delta\theta=2\xi$ induced by EWP 
contributions is seen to be rather small, of the order of $1.5^o$. Therefore, 
in practice these contributions can be neglected and the results of this analysis
are not likely to be of immediate relevance for an extraction of $\alpha$. 
However, we use this example to demonstrate that, in principle, 
the effects of EWP terms can be eliminated in a model-independent manner 
to allow a determination of the weak phase.

\subsection{Constraints on $\gamma$ from $B\to K\pi$ decays}

Recently the SU(3) relation (\ref{EW1}) between EWP contributions in 
$B^+\to K^0\pi^+$ and $B^+\to K^+\pi^0$ was obtained by Neubert and Rosner 
\cite{NR1}, and was used to derive information on $\gamma$ from the CP-averaged 
ratio
\bea\label{R*def}
R^{-1}_*=\frac{2[B(B^+\to K^+\pi^0) + B( B^-\to K^-\pi^0)]}
{B(B^+\to K^0\pi^+) + B(B^-\to \bar K^0\pi^-)}~.
\eea
Further constraints on the weak phase were shown to be provided by separate 
$B^+$ and $B^-$ branching ratio measurements if rescattering effects can be
neglected \cite{NR2}. 
In the present section we will review the arguments of 
\cite{NR1}, and then apply Eq.(\ref{EW2}), the second relation between EWP 
amplitudes in $B\to K\pi$, to the ratio \cite{FM}
\bea\label{Rdef}
R=\frac{B(B^0\to K^+\pi^-) + B(\bar B^0\to K^-\pi^+)}
{B(B^+\to K^0\pi^+) + B(B^-\to \bar K^0\pi^-)}~.
\eea
Our purpose here is to possibly eliminate uncertainties in $R$ due to EWP 
contributions in a model-independent manner. These contributions were argued 
in \cite{EWP,GR} to be color-suppressed and were calculated in 
specific model calculations \cite{FM,AKL} to be very small. Assuming that they 
can be neglected, and that the same applies to certain rescattering effects,
one obtains the bound \cite{FM} $R\ge \sin^2\gamma$ which can be useful provided
that $R<1$. Furthermore, measuring the CP asymmetry in $B \to K^{\pm}\pi^{\mp}$ 
would constrain $\gamma$ even if $R\ge 1$ \cite{GR}.
Here we will attempt to obtain a model-independent generalization of the bound 
$R\ge \sin^2\gamma$ including EWP effects 
\cite{FLEISCH}. The role of rescattering effects \cite{Rescat},
and possible limits on such effects \cite{GR,Limits}, were discussed 
elsewhere.
  
The amplitudes of the two decay processes appearing in $R^{-1}_*$ are given by 
\cite{GHLR,EWP}
\bea
\sqrt2 A(B^+\to K^+\pi^0) &=& -\lambda_u^{(s)}(T+C+P_{uc}+A) -
\lambda_t^{(s)}(P_{ct}-\sqrt2 P^{EW}(B^+\to K^+\pi^0))~,
\nonumber\\
A(B^+\to K^0\pi^+) &=& \lambda_u^{(s)}(P_{uc}+A) +
\lambda_t^{(s)}(P_{ct}+P^{EW}(B^+\to K^0\pi^+))~.
\eea
The contribution of the QCD penguin amplitude with an internal charm quark was 
included in $P_{uc}=P_u-P_c$ and $P_{ct}=P_t-P_c$ by making use of the 
unitarity of the CKM matrix. Using (\ref{EW1}), the first amplitude can be 
written as
\bea\label{K+pi0}
\sqrt2 A(B^+\to K^+\pi^0) &=& - |\lambda_u^{(s)}|(T+C)(e^{i\gamma}-\delta_{EW})
- \lambda_u^{(s)}(P_{uc}+A) - \lambda_t^{(s)}(P_{ct}+P^{EW})~,
\eea
where $P^{EW}\equiv P^{EW}(B^+\to K^0\pi^+)$ and
\bea\label{deltaEW}
\delta_{EW} = -\frac{3}{2}\left|\frac{\lambda^{(s)}_t}{\lambda^{(s)}_u}\right|
\frac{c_9+c_{10}}{c_1+c_2}~,
\eea
where $|\lambda^{(s)}_t/\lambda^{(s)}_u|=|V_{tb}V_{ts}/V_{ub}V_{us}|\approx
|V_{cb}/V_{us}V_{ub}|$.

Therefore,
\bea
& &R^{-1}_*=\nonumber\\
& &\qquad \frac{|\epsilon e^{i\phi_T}(e^{i\gamma}-\delta_{EW}) + 
\epsilon_A e^{i\phi_A}e^{i\gamma} - e^{i\phi_P}|^2 + 
|\epsilon e^{i\phi_T}(e^{-i\gamma}-\delta_{EW}) + 
\epsilon_A e^{i\phi_A}e^{-i\gamma} - e^{i\phi_P}|^2}
{|\epsilon_A e^{i\phi_A}e^{i\gamma}-e^{i\phi_P}|^2 + 
|\epsilon_A e^{i\phi_A}e^{-i\gamma}-e^{i\phi_P}|^2}~,
\eea
where we denote 
\bea
\epsilon e^{i\phi_T} &=& \frac{|\lambda_u^{(s)}|(T+C)}
{|\lambda_t^{(s)}||P_{ct}+P^{EW}|}~,
\qquad
\epsilon_A e^{i\phi_A} = \frac{|\lambda_u^{(s)}|(P_{uc}+A)}
{|\lambda_t^{(s)}||P_{ct}+P^{EW}|}~,
\eea
and $\phi_P = \mbox{Arg}(P_{ct}+P^{EW})$.
To first order in the small parameter $\epsilon\simeq 0.24$ \cite{NR1}, 
obtained through \cite{GRL}
\bea
\epsilon = \sqrt2 \frac{V_{us}}{V_{ud}}\frac{f_K}{f_\pi}
\frac{|A(B^+\to \pi^0\pi^+)|}{|A(B^+\to K^0\pi^+)|}~,
\eea
the ratio $R^{-1}_*$ is independent of the rescattering parameter $\epsilon_A$
and is given by
\bea\label{R*theor}
R^{-1}_* = 1 - 2\epsilon \cos\Delta\phi (\cos\gamma - \delta_{EW}) + 
{\cal O}(\epsilon^2)~,
\qquad
\Delta\phi=\phi_T-\phi_P~.
\eea
This then implies the bound \cite{NR1}
\bea\label{boundR*}
|\cos\gamma - \delta_{EW}| \ge \frac{|1-R^{-1}_*|}{2\epsilon}~,
\eea
which can set new constraints on $\gamma$ if $R_*\ne 1$. The central value of
a recent measurement \cite{CLEO}, $R_*=0.47\pm 0.24$, lies two standard
deviations away from one.

We now proceed to study the ratio $R$. Applying the relation (\ref{EW2}) to the 
corresponding EWP contributions, we find 
\bea
A(B^0\to K^+\pi^-) &=& -\lambda_u^{(s)} (T+P_{uc}) - \lambda_t^{(s)}
(P_{ct}+P^{EW})\nonumber\\
&+& \frac34\lambda_t^{(s)}
\left[ \frac{c_9-c_{10}}{c_1-c_2} (-T+C+A-E) -
\frac{c_9+c_{10}}{c_1+c_2} (-T-C+A+E) \right]\nonumber\\
&=&   -|\lambda_u^{(s)}|(T+P_{uc})
\left( e^{i\gamma} - \delta'_{EW} \right) - \lambda_t^{(s)} 
(P_{ct}+P^{EW})~,
\eea
where $P^{EW}$ is defined as in (\ref{K+pi0}), and $\delta'_{EW}$ (containing 
the EWP contribution) is defined by 
\bea
\delta'_{EW} &=& -\frac{3}{4}\left|\frac{\lambda^{(s)}_t}{\lambda^{(s)}_u}\right|
\left[\frac{c_9-c_{10}}{c_1-c_2}\cdot
\frac{-T+C+A-E}{T+P_{uc}} -
\frac{c_9+c_{10}}{c_1+c_2}\cdot \frac{-T-C+A+E}{T+P_{uc}}\right]\nonumber\\
\label{delta'EW}
 & &\simeq
-\frac{3}{2}\left|\frac{\lambda^{(s)}_t}{\lambda^{(s)}_u}\right|
\kappa \frac{C-E}{T+P_{uc}}~.
\eea
Here we made use of the approximate equality (\ref{approx}).

The ratio $R$ (\ref{Rdef}) can then be written as
\bea
R = 
\frac{|\epsilon' e^{i\phi'_T}(e^{i\gamma}-\delta'_{EW}) - e^{i\phi_P}|^2 + 
|\epsilon' e^{i\phi'_T}(e^{-i\gamma}-\delta'_{EW}) - e^{i\phi_P}|^2}
{|\epsilon_A e^{i\phi_A}e^{i\gamma}-e^{i\phi_P}|^2 + 
|\epsilon_A e^{i\phi_A}e^{-i\gamma}-e^{i\phi_P}|^2}~,
\eea
where 
\bea
\epsilon' e^{i\phi'_T} &=& \frac{|\lambda_u^{(s)}|(T+P_{uc})}
{|\lambda_t^{(s)}||P_{ct}+P^{EW}|}~.
\eea
Expanding again in powers of $\epsilon'$ and keeping only the linear terms, we 
obtain
\beq\label{Rtheor}
R = 1 - 2\epsilon'\cos(\Delta\phi'+\delta\phi)|\cos\gamma - \delta'_{EW}| 
+ {\cal O}(\epsilon'^2) + {\cal O}(\epsilon_A)~,
\eeq
where $\Delta\phi'=\phi'_T-\phi_P,~\delta\phi=$Arg$(\cos\gamma-\delta'_{EW})$.

Let us compare the structure of the two ratios $R$ (\ref{Rtheor}) and 
$R^{-1}_*$  (\ref{R*theor}) to first order in the small parameter 
$\epsilon'\approx\epsilon$. (These two parameters are equal up to corrections 
of order $|C/T|\simeq 0.2$ and $|P_{uc}/T|$). 
First, we note that $R$ depends on final state rescattering ($\epsilon_A$) 
whereas $R^{-1}_*$ is unaffected by such effects. This feature was already 
noted
in \cite{NR1}. The dependence of these ratios on EWP contributions is encoded
in  the parameters $\delta'_{EW}$ and $\delta_{EW}$. Whereas $\delta_{EW}$ 
(\ref{deltaEW}) is real and is given in terms of known Wilson coefficients 
and CKM factors, 
$\delta'_{EW}$ (\ref{delta'EW}) is in general complex and contains also the 
ratio $(C-E)/(T+P_{uc})$
depending on tree-level hadronic matrix elements. One usually assumes that
this ratio is smaller than one, given roughly by the color-suppression factor 
measured in $B\to \bar D\pi$ \cite{NS}. Thus 
\bea\label{EWratio}
|\delta'_{EW}/\delta_{EW}| \simeq |C/T| \simeq 0.2~.
\eea
Namely, EWP effects in $R$ are smaller than in $R^{-1}_*$ by a factor of about
5, in accord with \cite{EWP,GR}. A much smaller value than (\ref{EWratio}) 
was obtained in a model-dependent calculation \cite{FM}. 

Neglecting rescattering effects in $B^+\to K^0\pi^+$ \cite{Rescat,Limits}, 
(\ref{Rtheor}) implies the bound
\bea\label{boundR}
|\cos\gamma - \delta'_{EW}| \ge \frac{|1-R|}{2\epsilon'}~,
\eea
quite similar to (\ref{boundR*}).~~$\delta'_{EW}$ has a very small magnitude,
$|\delta'_{EW}|\simeq 0.2\delta_{EW}=0.13$, where we used $\delta_{EW}=0.63$
\cite{NR1}. Therefore, 
in spite of the uncertainty in the phase of $\delta'_{EW}$,
this constraint on $\gamma$ can potentially become useful provided that  
a value for $R$ is measured which is {\it different} from 1 (not necessarily 
smaller than 1 as required by \cite{FM}). 
For a given value of $|\delta'_{EW}|$, the allowed region for $\cos\gamma$ is
given by the constraint
\beq\label{cons1}
|\cos\gamma| > \frac{|1 - R|}{2\epsilon'} - |\delta'_{EW}|~,
\eeq
provided that
\beq\label{cons2}
1+|\delta'_{EW}| \ge \frac{|1-R|}{2\epsilon'} \ge |\delta'_{EW}|~.
\eeq
Eqs.~(\ref{cons1}) and (\ref{cons2}) exclude a region around $\cos\gamma=0$.
For $\epsilon'\simeq 0.24,~|\delta'_{EW}|=0.13$, this requires
$0.06 \le |1-R| \le 0.54$. The presently measured value of $R$, 
$R = 1.0 \pm 0.4$ \cite{CLEO}, largely overlaps with this region. 
We note that fixing the strong phase of $\delta'_{EW}$ by theoretical 
arguments can further sharpen these bounds.

It is possible to improve this constraint on $\gamma$ by combining the data on 
$R$ with a measurement of the pseudo-asymmetry $A_0$ \cite{GR}
\beq\label{A0def}
A_0=\frac{B(B^0\to K^+\pi^-) - B(\bar B^0\to K^-\pi^+)}
{B(B^+\to K^0\pi^+) + B(B^-\to \bar K^0 \pi^-)}~.
\eeq
One finds, to first order in $\epsilon'$,
\beq\label{A0theor}
A_0 = 
2\epsilon' \sin\gamma \sin\Delta\phi' + {\cal O}(\epsilon'^2) + 
{\cal O}(\epsilon_A)~.
\eeq
For $\delta'_{EW}=0$ and for given $\epsilon'$, $R$ (\ref{Rtheor}) and $A_0$
(\ref{A0theor}) determine $\gamma$ up to a fourfold ambiguity \cite{GR}.
In reality, since $|\delta'_{EW}|\simeq 0.13$ is very small, the solutions for
$\gamma$ are given by narrow bands corresponding to the uncertainty in the 
strong phase of $\delta'_{EW}$. 

\section{Conclusions}

Electroweak penguin amplitudes play an important role in various attempts to 
determine CKM phases from rate measurements. In $\Delta S=1$ $B$ decays
their contribution is comparable to that arising from the current-current 
terms in the weak Hamiltonian. It is therefore important to have an accurate 
theoretical control over their effect.  Based on flavor SU(3) and dominance of
$Q_9$ and $Q_{10}$ EWP operators, we presented a general method for 
relating the EWP contributions to tree-level amplitudes in $B$ decays to a
pair of charmless mesons. This reduces in a model-independent way the number 
of hadronic amplitudes parametrizing $B$ decays. SU(3) breaking effects on
these relations were studied in some cases in a model-dependent way and were 
found to be small \cite{NR1}. 

We applied these relations to three cases, a determination of $\alpha$ from
$B\to\pi\pi$ and two ways of constraining $\gamma$ from $B\to K\pi$ decays.
In the first case (where only isospin was used)
and when studying the ratio $R^{-1}_*$ in $B^{\pm}$ decays
\cite{NR1} (where SU(3) flavor was employed), constraints were obtained which 
were free of hadronic 
uncertainties. On the other hand, a study of $\gamma$ through the ratio $R$ 
in $B^{0,\pm}\to K\pi^{\pm}$ depends on the knowledge of the ratio of 
certain tree-level amplitudes. Neglecting rescattering effects, we used the 
smallness of this ratio to argue that useful constraints on $\gamma$ can be 
obtained from $R$ provided that $R$ is different from one. 

\acknowledgements
D.P. is grateful to Rich Lebed for a discussion of the results of the paper 
\cite{GrLe}. This work is supported by the National Science Foundation and by 
the United States - Israel Binational Science Foundation under Research 
Grant Agreement 94-00253/3.

\newpage
\appendix
\section{Four-quark operators with well-defined SU(3) transformation properties}

We give in this Appendix the four-quark operators appearing in the weak 
Hamiltonian for $b$ decays. They are defined as (in notation $\bar q_1\bar 
q_3 q_2\simeq (\bar bq_1)(\bar q_2 q_3)$)
\begin{itemize}
\item $\Delta S=+1$ operators
\bea
\overline {\bf 15}_{I=1} &=& -\frac12(\bar u\bar su+\bar s\bar uu) +
\frac12 (\bar d\bar sd+\bar s\bar dd)\\
\overline {\bf 15}_{I=0} &=& -\frac{1}{2\sqrt2}(\bar u\bar su+\bar s\bar uu) -
\frac{1}{2\sqrt2} (\bar d\bar sd+\bar s\bar dd) + \frac{1}{\sqrt2}\bar s\bar 
ss\\
{\bf 6}_{I=1} &=& -\frac12(\bar u\bar su-\bar s\bar uu)
+ \frac12 (\bar d\bar sd-\bar s\bar dd)\\
\overline {\bf 3}^{(a)}_{I=0} &=& -\frac12(\bar u\bar su-\bar s\bar uu)
- \frac12 (\bar d\bar sd-\bar s\bar dd)\\
\overline {\bf 3}^{(s)}_{I=0} &=& \frac{1}{2\sqrt2}(\bar u\bar su+\bar s\bar uu) +
\frac{1}{2\sqrt2} (\bar d\bar sd+\bar s\bar dd) + \frac{1}{\sqrt2}\bar s\bar ss\,.
\eea

\item $\Delta S=0$ operators
\bea
\overline {\bf 15}_{I=\frac32} &=& -\frac{1}{\sqrt3}(\bar u\bar du+\bar d\bar uu) +
\frac{1}{\sqrt3} \bar d\bar dd\\
\overline {\bf 15}_{I=\frac12} &=& -\frac{1}{2\sqrt6}(\bar u\bar du+\bar d\bar uu) +
\frac12\sqrt{\frac32}
(\bar s\bar ds+\bar d\bar ss) - \frac{1}{\sqrt6}\bar d\bar dd\\
{\bf 6}_{I=\frac12} &=& \frac12(\bar d\bar ss-\bar s\bar ds)
+ \frac12 (\bar u\bar du-\bar d\bar uu)\\
\overline {\bf 3}^{(a)}_{I=\frac12} &=& -\frac12(\bar u\bar du-\bar d\bar uu)
+ \frac12 (\bar d\bar ss-\bar s\bar ds)\\
\overline {\bf 3}^{(s)}_{I=\frac12} &=& \frac{1}{2\sqrt2}(\bar u\bar du+\bar d\bar uu) +
\frac{1}{2\sqrt2} (\bar d\bar ss+\bar s\bar ds) + \frac{1}{\sqrt2}\bar d\bar dd\,.
\eea
\end{itemize}

We also list the relative normalization between SU(3) reduced matrix elements 
used in this paper and in \cite{GHLR}:
\beq
a_1 = -\frac{1}{\sqrt3}\{1\},~~a_2 = -2\sqrt{\frac23}\{8_1\},~~a_3 = 
-\frac{2}{\sqrt3}\{8_2\},~~a_4 = \frac{4}{\sqrt5}\{8_3\},~~a_5 = 
3\sqrt{\frac65}\{27\}~.
\eeq

\thispagestyle{plain}
\begin{figure}[hhh]
 \begin{center}
 \mbox{\epsfig{file=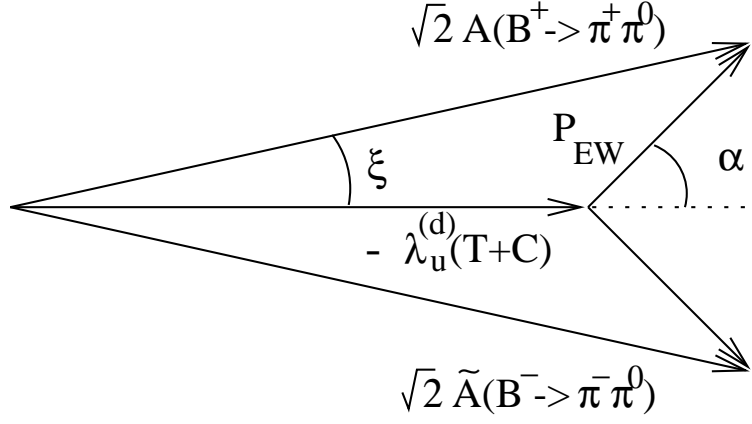,width=10cm}}
 \end{center}
 \caption{EW penguin effects in the decay amplitude $A(B^+\to\pi^0\pi^+)$ and its
charge conjugate $\tilde A(B^-\to\pi^0\pi^-)\equiv e^{2i\gamma}A(B^-\to\pi^0\pi^-)$.}
\label{fig1}
\end{figure}

\end{document}